# A Traffic-Aware Medium Access Control Mechanism for Energy-Efficient Wireless Network-on-Chip Architectures

Naseef Mansoor[1], Abhishek Vashist[2], M Meraj Ahmed[2], Md Shahriar Shamim[4], Syed Ashraf Mamun[3], Amlan Ganguly[2]

**Abstract**—Wireless interconnection has emerged as an energy efficient solution to the challenges of multi-hop communication over the wireline paths in conventional Networks-on-Chips (NoCs). However, to ensure the full benefits of this novel interconnect technology, design of simple, fair and efficient Medium Access Control (MAC) mechanism to grant access to the on-chip wireless communication channel is needed. Moreover, to adapt to the varying traffic demands from the applications running on a multicore environment, MAC mechanisms should dynamically adjust the transmission slots of the wireless interfaces (WIs). Such dynamic adjustment in transmission slots will result in improving the utilization of the wireless medium in a Wireless NoC (WiNoC). In this paper we present the design of two dynamic MAC mechanisms that adjust the transmission slots of the WIs based on predicted traffic demands and allow partial packet transfer. Through system level simulations, we demonstrate that the traffic aware MAC mechanisms are more energy efficient as well as capable of sustaining higher data bandwidth in WiNoCs.

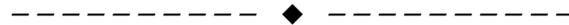

## 1 INTRODUCTION

Network-on-Chip (NoC) has emerged as the enabling technology for catering to the communication needs of high performance applications in modern multi-core System-on-Chips (SoCs) [1]. NoCs enable transfer of on-chip data using wormhole switching where packets are broken into smaller units called flits that can be transmitted in a single cycle over pipelined links and switch stages [2]. Among different alternatives, Mesh based NoC architecture is primarily used in multicore products [3] as it is easy to design, debug, and manufacture. Despite its advantages, due to the multi-hop communication over the metal interconnects, traditional mesh based NoC architectures are not scalable in terms of performance and energy consumption. Long range metal wires in a mesh based NoC [4] and ultra-low-latency and low-power express channels between communicating cores [5] have been proposed to solve these inefficiencies. However, the performance gain of these approaches are limited due to metal/dielectric based interconnection. This has led to the exploration of novel interconnect technologies like on-chip photonic interconnect [6], multi-band RF transmission line interconnects (RFI) [7], and wireless interconnects [8]. Among, these alternatives, wireless interconnect operating in millimeter wave (mm-wave) band is nearer term solution due to the compatibility of the enabling technology of miniature metallic antennas and transceivers [9] with CMOS fabrication processes. Such mm-wave wireless NoC (WiNoC) architectures are shown to improve the performance and the energy efficiency in on-chip data transfer significantly [9]. In mm-wave WiNoCs the wireless channel is shared between multiple wireless transceivers. Moreover, to enable communication between the transceivers oriented in different directions with respect to each other in the WiNoC, non-directional antennas are proposed [10]. However, utilizing the full potential of the novel mm-wave interconnect technology in a WiNoC requires overcoming two critical design challenges: i) design of efficient, simple and fair medium access control (MAC) mechanism for the shared wireless medium and ii) allocation of the shared wireless bandwidth effectively. Due to the distributed and low-overhead implementation, and fairness in channel access, a token passing based Time Division Multiple Access (T-MAC) is used in many WiNoC architectures [11], [12], [13].

In a token based MAC, a single WI possessing the token gains access to the wireless medium to transmit for a certain number of time slots. Therefore, the T-MAC is a distributed, and asynchronous wireless channel access mechanism. The MAC should also manage the sharing of the wireless communication medium depending on traffic variation in the WiNoC in order to maximize performance. In a NoC the traffic demand through the switches vary both temporally and spatially depending on the application [14],[15]. The traffic pattern in multicore chips is characterized as bursty and self-similar [16]. This causes sudden and large variations in the traffic demands on the NoC. Therefore, the MAC for a WiNoC should be able to dynamically

1. *Naseef Mansoor is with the Department of Electrical and Computer Engineering and Technology, Minnesota State University, Mankato, MN, 56001. E-mail: naseef.mansoor@mnsu.edu*
2. *Avishek Vashist, M Meraj Ahmed and Amlan Ganguly are with the Department of Computer Engineering, Rochester Institute of Technology, Rochester, NY. 14623, E-mail: (av8911, ma9205, axgeec)@rit.edu.*
3. *Syed Asraf Mamun is with the Thomas B. Golisano College of Computing and Information Science, Rochester Institute of Technology, Rochester, NY. 14623. E-mail: sam7753@rit.edu.*
4. *Md Shahriar Shamim is with the Failure Analysis and Fault Isolation team in Intel Corp., Portland, OR. Email: Shahriar.shamim@intel.com*

allocate transmission slots to WIs in response to sudden and large variations in traffic. Moreover, in T-MAC the token is released to the next WI after the transmissions. Therefore, it is necessary to transmit whole packets to maintain the integrity of the wormhole switching protocol [17]. If body flits are transmitted without header address information, it will be impossible to associate them with the correct packet at the receiving WIs. Therefore, partial packets can not be transmitted to correct destination WIs because the wireless medium is shared among all the WIs unlike point-to-point wired links. This results in inefficient utilization of the wireless channel limiting the performance benefit of the wireless interconnect. Moreover, for large packets, this requires large buffers in the Virtual Channels (VCs) of the WI to store entire packets before they can be transmitted over the wireless medium. Therefore the MAC mechanism should allow partial packet transfer while maintaing the integrity of wormhole switching in the wirelss interconnects.

In this work, we propose two TDMA based dynamic MAC mechanisms; Proportional Slot Allocation Mechanism (P-SAM) MAC and Demanded Slot Allocation Mechanism (D-SAM) MAC that are able to track traffic demands and dynamically allocate the transmission slot durations to WIs while allowing partial packet transfer. The proposed slot allocation mechanisms rely on the tracking mechanism to predict the traffic demands for the WIs rather than being based on current utilization of links or swtiches. This minimizes reaction times to sudden and bursty variation in traffic. The specific contributions of this paper are:

1. Design of a low complexity and accurate traffic tracking mechanism that is able to predict the traffic demand of a WI.

2. Propose two dynamic MAC mechanisms: P-SAM MAC and D-SAM MAC that are able to adjust the slot durations based on the predictions.

3. Propose an appropriate wormhole switching mechanism for the dynamic MACs allowing partial packet transmission.

4. Evaluate and compare the proposed MAC mechanism with different state-of-the-art MAC mechanisms.

The rest of the paper is organized as follows: section 2 discusses most recent and relevant related works, section 3 describes the dynamic MAC mechanisms. Section 4 presents the evaluations of the proposed MAC mechanism and finally section 5 concludes the paper.

## 2 RELATED WORKS

The design of the MAC mechanism enabling collision free and efficient utilization of the wireless channel in an on-chip environment is constrained by area, power and buffer overheads. Due to these restrictions, complex MAC mechanisms used in conventional networks are not suitable for WiNoCs [18]. Hence, design of efficient, low overhead, and fair MAC mechanisms are considered as one of the critical challenges for WiNoCs [19]. A synchronous and distributed MAC mechanism (SD-MAC) is proposed in [20] for the Ultra-Wide-Band (UWB) WiNoC architectures. Such WiNoC uses impulse based transceivers where the communication range is a few millimeters. Furthermore, in order to access the wireless channel, the WIs share the request control packets over wired links. Thus, such MAC mechanism cannot be adopted for WiNoCs where the WIs is more than a millimeter apart. A hybrid MAC mechanism combining both TDMA and FDMA based access is proposed for CNT based WiNoC architectures in [8]. Although, the CNT based antennas enables communication among the WIs distributed over the chip, their implementation in current CMOS process are challenging. Authors in [21], [22] has proposed an mm-wave WiNoC architecture with multiple non-overlapping channels to enable FDMA based medium access. However, such FDMA based approach is non-trivial from the persperctive of transceiver design and the number of concurrent channels are not easily scalable. WiNoC architecture with CDMA based MAC mechanism proposed in [23] also enables concurrent communication among the WIs and efficiently utilizes the wireless bandwidth. However, CDMA requires coherent Binary Phase Shift Keying (BPSK) receiver along with Analog-to-Digital Convereters (ADC) in the transceivers making the design significantly challenging. Similar to the CDMA, a distributed MAC with strictly synchronized orthogonal request packets is proposed in [18]. In [24], authors proposed a TDMA based CSMA MAC mechanism for WiNoC architectures. However, the CSMA MAC mechanism suffers from performance degradation at high traffic loads due to back-off based collision recovery as demonstrated in [25]. Therefore, a simple, distributed and low-overhead token passing based MAC (T-MAC) mechanism is adopted for many WiNoC architectures [11], [12]. In the T-MAC mechanism, the access to wireless medium is granted to a WI by the possession of a token, circulating among the WIs, organized in a virtual ring. However, the T-MAC mechanism is unaware of the varying traffic demand through the WIs, resulting in inefficient utilization of the energy efficient wireless channel. A dynamic radio access control mechanism (RACM) proposed in [26] shows that dynamic allocation of slots to the WIs improves the performance of a WiNoC. In the RACM MAC mechanism the unused slot by the WIs in an epoch is redistributed among the WIs in the next epoch based on their current slot usage. A similar MAC mechanism reported in [27] allocates slot to the wireless transceivers based on the link utilization. Although these dynamic MAC mechanisms are novel and provides promising solution for traffic adaptility in WiNoC architectures, they are reactive with slow system level response as they depend only on the utilization of the WIs based on current traffic behavior. Moreover, these MAC mechanisms uses packet-switching for maintain the integrity of the wormhole switching in the wireless interconnect. Such mechanism results in inefficient utilization of the wireless channel while requiring large buffers in the VCs of the WIs to store large packets before they can be transmitted over the wireless medium. In this work, we propose the design of a dynamic MAC mechanism that can proactively allocate slots to the WIs based on their traffic demands while allowing partial packet transfer.

# 3 DYNAMIC MAC MECHANISM FOR WINOC

In this section, we discuss the proposed dynamic MAC mechanisms. We track the variations in traffic to predict the traffic demand for each WI. The MACs then allocate the access of the shared wireless medium to the WIs based on the predictions.

## 3.1 Tracking and Prediction of Traffic Demand

In the proposed MAC mechanism, a Proportional-Integral-Derivative (PID) technique is used to track the traffic demand of a WI. We define the traffic demand as the number of flits that the WI needs to transmit over a window of time, which we will refer to as an *epoch*. This tracking is used to predict the future traffic demand at each WI for the slot allocation mechanisms. An accurate prediction mechanism is required for a pro-active transmission slot allocation instead of reacting to only current traffic demand. For the choice of the tracking mechanism, we focus on both implementational complexity (i.e. area overhead and computational complexity) and accuracy. Complex tracking and prediction mechanisms for systems with complex random behaviors, such as Kalman Filtering [28], Artificial Neural Network (ANN) [29] and Support Vector Regression (SVR) [30] suffers from high implementational complexity due to the sophisticated nature of these algorithms. On the other hand, a PID based tracking mechanism is widely used in control systems for its simplicity, robustness and efficiency [31]. The proportional and the integral part (i.e. first and second term) in equation (1) captures current and steady state demand of the WI respectively, whereas the differential part (i.e. the third term) captures the temporal variation in demand. Hence, such PID based prediction mechanism captures the change in WI demand due to both the short term and long term traffic variation, suitable for bursty traffic in NoCs. Due to these advantages of the PID tracking, we propose to use this mechanism for predicting the traffic demand of the WIs in the next epoch. According to PID tracking the predicted traffic demand of a WI *i*, $D_j^{i^{pred}}$ for epoch *j* is given by,

$$D_j^{i^{pred}} = K_p * D_{j-1}^{i^{Act}} + K_I * \overline{D^i} + K_D * (D_{j-1}^{i^{Act}} - D_{j-2}^{i^{Act}}) . \quad (1)$$

Where, $D_{j-1}^{i^{Act}}$ and $D_{j-2}^{i^{Act}}$ are the actual traffic demand of the WI *i* for epoch *j-1* and *j-2* respectively. $\overline{D^i}$ is the average traffic demand of the WI from epoch 0 to *j-2*. The $K_p$, $K_I$ and $K_D$ are weights that minimizes the error in the predicted traffic demand with respect to the actual demand. The methodology to optimize the value of the weights $K_p$, $K_I$ and $K_D$ for minimizing the prediction error is discussed in subsection 4.4 after explaining the simulation environment used to compute the error.

The WIs are equipped with a Prediction Unit (PU) for calculating the predicted traffic demand according to (1). The PU consists of two counters **Epoch_counter**, and **Demand_counter** and three registers **Demand<sub>average</sub>**, **Demand<sub>previous</sub>** and **Demand<sub>self</sub>**. The **Epoch_counter** counts the duration of an epoch and at the end of the current epoch, it is set to the number of slots (i.e. fixed or variable) in the next epoch. The **Demand_counter** is used to compute the actual traffic demand of a WI in an epoch (i.e. $D_{j-1}^{i^{Act}}$

in (1)) and is incremented when a flit is routed to the wireless port. The registers **Demand<sub>average</sub>** and **Demand<sub>previous</sub>** representing the terms $\overline{D^i}$ and $D_{j-2}^{i^{Act}}$ in (1) respectively, stores the average and previous actual traffic demand of the WI from the previous epoch. When the **Epoch_counter** expires, the predicted traffic demand of a WI is calculated using the **Demand_counter**, **Demand<sub>average</sub>** and **Demand<sub>previous</sub>**. The predicted traffic demand is stored in **Demand<sub>self</sub>** (i.e. $D_j^{i^{pred}}$ in (1)). The **Demand<sub>average</sub>** is then updated by the average of the values in **Demand_counter** and **Demand<sub>average</sub>**. The **Demand<sub>previous</sub>** is updated using the **Demand_counter** to capture the actual traffic demand from the previous epoch. The predicted traffic demand of a WI is shared with other WIs as it is required for the allocation of slot durations at all the WIs as can be seen in the next subsection. The mechanism to share this information using a *slot information packet* is discussed in subsection 3.3.

## 3.2 Slot Allocation Mechanism

The slot allocation mechanism is responsible for allocating the slot duration. The allocated slot duration is equal to the number of flits the WI is allowed to transmit over the wireless medium before the right of medium access passes to the next WI. For the proposed dynamic MAC mechansism, the slot duration allocation is based on the predicted traffic demand of the WIs. Consequently, the slot durations are adjusted to cope up with the traffic variation. We propose two slot allocation mechanisms, proportional slot allocation mechanism (P-SAM) and demanded slot allocation mechanism (D-SAM) for determining the slot duration of the WIs based on the predictive demand. The WIs are equipped with an allocation unit that implements one of the two slot allocation mechanisms.

### 3.2.1 Proportional Slot Allocation (P-SAM)

In a basic token based TDMA MAC for WiNoC [11] each WI possesses the token for a fixed maximum duration before releasing it to the next WI for medium access. This neglects the unequal traffic demands at the WIs and their temporal variations by permitting WIs to transmit depending only on their current traffic demand without knowledge of the demand of others which maybe more than its own. In the P-SAM scheme, the slot duration of a WI is dynamically adapted based on the proportion of the predicted traffic demand of the WI relative to other WIs. To make this proportional slot duration allocation, a fixed epoch duration similar to the token based MAC is assumed which is distributed among the WIs depending on their relative predicted traffic demands. The number of transmission slots for a WI in an epoch is allocated dynamically at the start of each epoch. The allocated slot duration of a WI *i*, at epoch *j+1*, $S_{j+1}^i$, is given by

$$S_{j+1}^i = \frac{D_j^{i^{pred}}}{\sum_{i=1}^N D_j^{i^{pred}}} \times E_F . \quad (2)$$

Here, $D_j^{i^{pred}}$ is the predicted traffic demand for WI *i* for epoch *j* calculated using (1), $E_F$ is the constant epoch duration in terms of the number of flits that can be transmitied over the wireless medium in an epoch and *N* is the number of WIs in the system. Hence, in the P-SAM scheme the duration of epoch remains constant. However, each individual transmission slot for a WI within an epoch changes among epochs based on the predicted demand among the WIs.

To enable this P-SAM scheme, the Allocation Unit (AU) contains a **Slot_counter** and a register file **REG_demand**. The **Slot_counter** is used to count the slot duration for a WI (the duration for which the WI will have access to the wireless medium) in an epoch. The register file, **REG_demand** is used to store the predicted demand of other WIs received from the *slot information packet*. In an epoch, when all the values of the **REG_demand** is updated and **Demand_self** is shared with other WIs, the slot duration for the WIs are calculated using the values in **Demand_self** and **REG_demand**. This value is then used to update the S**lot_counter** at the beginning of the next epoch. The **Epoch_counter** is set to the fixed epoch duration at the end of each epoch.

### 3.2.2 Demanded Slot Allocation (D-SAM)

It can be observed that the P-SAM proportionately distributes a fixed epoch duration among the WIs. However, if all WIs collectively have fewer flits to transmit than the number of flits which can be transmitted in an entire epoch, the fixed epoch duration will result in wasted transmission slots. Therefore, we explore the next allocation mechanism with dynamic epoch duration while still enabling partial packet transfer instead of sending whole packets on the fly as they arrive at the WIs. In the D-SAM scheme the slot duration is allocated dynamically based on the predicted traffic demand of the WI. However, unlike the P-SAM scheme, the maximum number of flits that can be transmitted in a slot is directly equal to the predicted demand of a WI. Hence, the total number of flits that can be transmitted in an epoch or the epoch duration, can vary among epochs for the D-SAM scheme. The allocated slot duration for WI, $i$ at epoch, $j+1$, $S_{j+1}^i$ is given by

$$S_{j+1}^i = D_j^{i_{pred}}. \qquad (3)$$

Where, $D_j^{i_{pred}}$ is the predicted traffic demand for WI $i$ at epoch $j$. Similar to the P-SAM scheme, in the D-SAM scheme the demand information shared using the *slot information packet*. This demand information is used to calculate the number of flits that can be transmitted in an epoch. The flits that can be transmitted in epoch $j+1$, $E_F^{j+1}$, is given by

$$E_F^{j+1} = \sum_{i=1}^{N} S_{j+1}^i . \qquad (4)$$

Here, $S_{j+1}^i$ is the allocated slot duration in the epoch $j+1$ for WI $i$. The duration of the next epoch is calculated after all the predicted demand information of the WIs are shared in an epoch. The allocation unit in a D-SAM contains the **Slot_counter** and the register file **REG_demand** as in the P-SAM MAC. However, unlike the P-SAM scheme, the **Epoch_counter** for the D-SAM scheme is updated with the sum of **REG_demand** and **Demand_self** at the end of each epoch.

In both P-SAM and D-SAM scheme the allocation of slot duration for a WI is adjusted dynamically at every epoch based on the demand of the WIs. We eliminate the possibility of starvation of WIs in accessing the wireless channel as the predicted traffic demand is computed for each epoch and all WIs with packets to send over the wireless medium is allocated a non-zero number of transmission slots. The difference in slot allocation between the proposed MAC mechanisms are shown via timing diagrams in Fig. 1.

### 3.3 Wireless Flow Control Mechanism

The wireless flow control mechanism enables an energy-efficient communication and partial packet transmission through the wireless medium similar to conventional wireline NoCs with virtual channel based routers. This results in lowering the buffer requirement in the output VCs of the WIs. Consequently, the energy consumption of the WIs are reduced. The partial packet contains one or more flits with same packet number. However, to ensure the correct flow of the partial packets, a WI needs to share the information about the partial packets it is going to transmit with other WIs before transmission begins. Thus, each WI generates a *slot information packet* at the beginning of its transmission with the information regarding the partial packets to be transmitted. Then this *slot information packet* is broadcast using the wireless medium to share the partial packet information with other WIs. The *slot information packet* consists of several fields such as a **Header, Size, Demand, ID and** several 3-tuples. The number of 3-tuples in a *slot information packet* is limited by the number of output VCs of the transmitting WI. The structure of the *slot information packet* is shown in Fig. 1. The **Header** is used to identify and differentiate the *slot information packet* from data packets. The

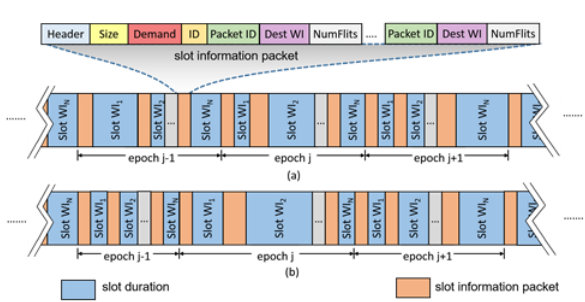

Fig. 1. Timing diagram of (a) P-SAM MAC and (b) D-SAM MAC.

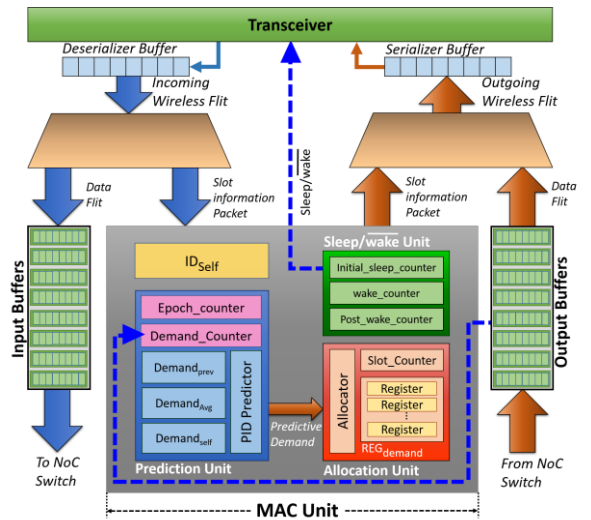

Fig. 2. Architecture of the WI with the proposed MAC unit.

*size* indicates the number of flits in the *slot information packet*. The maximum size of the *slot information packet* depends on the number of tuples in the *slot information packet* and is much smaller than a data packet. The **Demand** is populated with the value of the register **Demand**$_{self}$. In this way the predicted traffic demand of the WI required by the allocation schemes is shared with other WIs using the *Demand* field of the *slot information packet*. The identity of the WI is stored in the **ID**$_{self}$ register and is used to populate the *ID* of the *slot information packet* by the transmitting WI. This *ID* is used to index into the **REG**$_{demand}$ by the receiving WIs and the predicted demand value is updated with the *Demand*. The 3-tuples: (**DestWI**, **PktID**, **NumFlits**) in the *slot information packet* is used to enable partial packet transmission and correct routing. The 3-tuple containing information about the number of flits (i.e **NumFlits**) to be transmitted from the source WI to the destination (i.e., **DestWI**) along with the packet ID (i.e., **PktID**) of the packet the flits belong to. The **PktID** is used by the destination WIs to identify the appropriate VC to receive these flits. In case the **PktID** does not exist at the destination WI, the WI reserves an unoccupied VC for the new packet. Furthermore, to support broadcast traffic patterns, we propose to use a reserved value for the **DestWI** to denote a broadcast packet that is destined to all the WIs. This approach for generating broadcast addresses is similar to Internet Protocol (IP) [32].

Energy-efficiency in wireless communication is further achieved by using sleep transistors in the transceivers. We adopt the design of such sleepy transceivers from [33]. Unlike [33], we use the information in the *slot information packet* to generate the sleep/wake signals instead of global signaling wires to eliminate the routing complexity of the global lines. A sleep/wake unit in the WI turns the receivers on only when it is scheduled to receive flits in a transmission slot. Otherwise, the receiver circuit is turned off. The sleep/wake unit consists of three counters: **initial_sleep_counter**, **wake_counter** and **post_wake_counter**. These counters are set using the 3-tuples in the *slot information packet*. The **initial_sleep_counter** and **post_wake_counter** is set to the number of flits transmittied to other WIs before and after sending flits to this WI respectively. The **wake_counter** is set to the duration the receiver should be on, to receive the flits destined to the WI. The **initial_sleep_counter** starts at the beginning of each transmission slot. On expiration, the receiver is turned on and the **wake_counter** is started. The expiration of the **wake_counter** turns off the receiver and starts the **post_wake_counter**. The expirations of the **post_wake_counter** signifies the end of a transmission slot. At this point the receiver is turned on to receive the *slot information packet* for the next transmission slot. The transmitters on the other hand are always-off except its own transmission slot. Once the WI finished the transmission of the flits as per the *slot information packet*, the next WI automatically wakes up and creates and transmits its own *slot information packet* to signifiy the beginning of its transmission slot. The maximum duration of this transmission slot is computed according to either P-SAM or D-SAM. To maintain the transmission order of the WIs in an epoch, they are organized into a virtual ring and the unique value in the **ID**$_{self}$ register denotes the order of a WI in the ring.

The proposed MACs naturally adapt to the scenario where no WI is predicted to have any flits to transmit in a particular epoch. In case none of the WIs have any flits to send, the control *slot information packet* is continiuously shared with the other WIs. This allows for immediate transmission whenever any WI has a flit to send without any wake-up latency. However, similar to CSMA, this requires all the receivers in the WIs to be always-on until they receive specific sleep slot instructions from any WI. The architecture of a WI for the proposed dynamic MAC mechanism including the prediction unit, allocation unit and the sleep/wake unit is shown in Fig. 2. In the next section we evaluate the benefits of these MAC mechanisms in a WiNoC.

# 4 EXPERIMENTAL RESULTS

In this section, we evaluate the performance and energy efficiency of the proposed dynamic MAC based WiNoC architectures. Although, many different WiNoC architectures has been proposed in literature, for this work we have considered a Mesh based hybrid WiNoC architecture (i.e. WiMesh) with both wired and wireless links as a test case. The WiMesh architecture is discussed in details in the next subsection. The performance of the WiNoC architecture is measured as the peak achievable bandwidth per core or bandwidth and packet latency. The bandwidth of a WiNoC is determined as the average number of bits successfully routed to the destination cores per second from each source core. The packet latency is the average number of clock cycles required to successfully transmit a packet to the destination core. The energy efficiency of the WiNoC is measured as the packet energy which is the average energy (both dynamic and static) required to route a whole packet from source to destination through the NoC components (i.e. switches and the links). In the next subsection we describe the baseline WiMesh architecture used in this paper for evaluations.

## 4.1 WiMesh Architecture for the Test-Bed

We adopt a Mesh based WiNoC architecture as a test-bed for evaluating the proposed MACs. In the WiMesh architecture, each core is connected to a NoC switch using a wireline link. The switches are then connected with other switches in its cardinal directions (i.e. NSEW) using wireline interconnects to form a regular Mesh. The Mesh is chosen as it is a conventional NoC topology used in several multicore based products [3] and is relatively easy to design, verify, and manufacture. To provide single-hop shortcuts among the distant NoC switches to reduce the data transfer over multihop wireline paths, the wireless interconnects are overlaid on top of this Mesh topology by deploying the WIs at some of the NoC switches. To deploy the WIs for best performance gains, we adopt the optimization method outlined in subsection 4.3.

To realize the wireless interconnect, each WI contains on-chip antenna, transceiver circuit and serializer/deserializer buffers. As the WIs can be potentially at different angles with respect to each-other's axes, the radiation pattern for the antenna should be non-directional. The antenna should also provide the best power gain for the smallest area overhead. A CMOS compatible metal mm-wave zig-zag antenna operating at 60GHz mm-wave bands and bandwidth of 16GHz has been demonstrated to possess these characteristics [9]. Hence, we equip the WIs with such miniature on-chip zig-zag antenna to enable the long-range shortcuts among the WIs located at different part of the chip. On the other hand, to ensure high performance and energy-efficiency, the transceiver circuit has to provide a very wide bandwidth and consume low power. We adopt the transceiver design from [34], [35], [36] where low power design considerations are taken into account. Non-coherent on-off keying (OOK) modulation is chosen, as it allows relatively simple and low-power circuit implementation. The power and bandwidth of the OOK transceivers are adopted from fabricated prototypes demonstrated in 65nm technology [34], [35], [36]. The wireless transceiver is shown to dissipate 2.06pJ/bit sustaining a data rate of 16Gbps with a bit-error rate (BER) of less than $10^{-12}$ while occupying an area of $0.17mm^2$ in post-layout design using TSMC 65nm CMOS process. The serializer/ deserializer buffers realized through shift registers work as a data interface between the transceiver and NoC switch.

The WiMesh contains both short (e.g. wired interconnects) and long (e.g. wireless) links. In [17], a shortest path based routing is used to optimize the performance of such networks. Hence, we adopt such shortest path based routing scheme for the WiMesh architecture. We use a forwarding-table based routing through pre-computed shortest paths. The shortest path between any two pairs of nodes in the network is determined using a minimum spanning tree formed by Dijkstra's algorithm. The minimum spanning tree formed by the Dijkstra's algorithm depends on the chosen start node but the length of paths between any particular pair is independent of the start node. Hence, the minimum spanning tree is selected randomly. Furthermore, deadlock is avoided by transferring flits along the shortest path routing tree extracted by Dijkstra's algorithm, as it is inherently free of cyclic dependencies. Hence, each switch only has local forwarding information eliminating the need for maintaining non-scalable global routing information resulting in a scalable routing mechanism.

## 4.2 Simulation Platform

The NoC architectures (i.e. topology, switch architecture, flow control and routing mechanism, MAC scheme, link delay and bandwidth etc.) are characterized using a cycle accurate NoC simulator. The simulator accurately models the progression of the flits over the switches and links per cycle accounting for those flits that reach the destination as well as those that are stalled. The post-synthesis delay and the energy dissipation of the NoC components considering both dynamic and static power consumption are annotated into the simulator for evaluating the performance and energy efficiency of the NoC architectures. We consider a system size of 64 cores that represents the current trends in multicore chip design in the industry [37]. We also demonstrate the scalability of the proposed MAC by evaluating it for higher system size of 256 cores. In each experiment

| | Prediction Unit | Allocation Unit | | sleep/ wake Unit | Total | |
| --- | --- | --- | --- | --- | --- | --- |
| | | P-SAM | D-SAM | | P-SAM | D-SAM |
| Power (mW) | 0.1482 | 0.163 | 0.0759 | 0.062 | 0.373 | 0.286 |
| Area (um²) | 958.679 | 1264.32 | 373.319 | 406.05 | 2629.02 | 1738.048 |
| Delay (ns) | 0.12 | 0.12 | 0.13 | 0.14 | 0.14 | 0.14 |

with synthetic traffic, ten thousand iterations were performed eliminating transients in the first thousand iterations. We also evaluate the proposed MAC mechanism for application specific traffic scenarios. For the wired switches, we adopted a wormhole based flow control mechanism, where the packets are broken in smaller flow control units or flits [2]. The NoC switch is adopted from a three-stage pipelined design [38]. Each switch is considered to have 4 VCs with a buffer depth of 2. However, as the WIs handle a large volume of traffic, an increased number of VC of 8 with buffer depth of 16 is used. A moderate packet size of 64 flits is considered for all our experiments. The width of all wired links is considered to be same as the flit size, which is considered to be 32 bits. For the comparative study among the MAC mechanisms, we consider a wireless token passing based MAC mechanism (e.g. T-MAC) as the baseline MAC mechanism. In the baseline T-MAC mechanism, a token in the form of a wireless flit is circulated among the WIs in a round-robin fashion. A WI can only transmit one entire packet when it possesses the token to maintain the integrity of the wormhole routing mechanism over the wireless channel. Therefore the fixed duration of transmission slot in the baseline T-MAC is considered to be the size of one complete packet. This necessitates increased buffer depth of 64 flits to accommodate complete packets in the VCs of the WIs with this MAC.

The packet energy is estimated by adding the energy consumption in link (both wired and wireless) and switch traversals by the packet. The energy dissipation and delay of the wired link is obtained through Cadence simulations taking into account the specific lengths of each link based on the established topology in the 20mmx20mm die. For the wireless interconnect, we adopted the antenna [9] and the transceiver design from [34], [35], [36] as mentioned in the previous subsection. The NoC switches and proposed dynamic MAC units are synthesized from a RTL level design using 65nm standard cell libraries from TSMC using Synopsys. A 2.5GHz clock and 1V $V_{dd}$ representing the nominal frequency and voltage for the 65nm technology node is used for synthesis. The synthesis result of the proposed dynamic MAC unit is shown in Table 1. The power consumption and the delay of the MAC unit is considered in our simulations. However, as is evident from Fig. 2 the MAC unit is parallel to the datapath of flit transmission and reception. Therfore, its latency does not impact the overall data latency. However, the energy and delay overheads of circulating the *slot information packet* is considered in our simulations. The delay overhead due to the *slot information packet* depends on its size. In the WiMesh architecture, the maximum size of the *slot information packet* turns out to be 4 flits. Next, we use this simulation platform to optimize the WiMesh architecture for best performance before performing the evaluations of the proposed MAC mechanism.

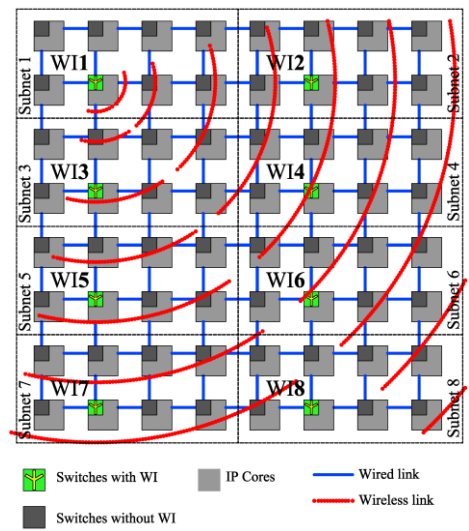

Fig. 3. 64 core WiMesh architecture with 8 WIs.

### 4.3 Optimizing the Baseline WiMesh

To optimize the baseline WiMesh architecture for best performance, we divide the Mesh in small logical subnets of equal size where the size denotes the number of cores in a subnet. The WIs are then deployed in one of the central switch of the subnet. A conceptual view of the WiMesh architecture is depicted in Fig. 3. The performance of the baseline WiMesh architecture (i.e. WiMesh with T-MAC) varies with different subnet sizes as the number of WIs in the system varies due to the WI deployment strategy. In the WiMesh architecture when the subnet size is large (i.e. the number of WIs is low), each WI is shared by many cores. This results in a high traffic load through the WIs as they provide long-range shortcuts between distant switches. On the other hand, when the subnet size is small (i.e. number of WIs is high), the interval between two consecutive channel accesses by a WI is large. Due to this long interval in channel access, the flits at the WIs waits longer. Hence, to achieve the best performance in the baseline WiMesh, the subnet size or number of WIs should be optimized.

Fig. 4 depicts the peak achievable bandwidth per core for the 64-core baseline WiMesh with different subnet sizes for uniform-random traffic pattern. We follow the same subnet based WI deployment strategy while generating the WiMesh configurations with different number of subnets. Uniform-random traffic pattern is used for this optimization to capture the effect of both short and long distance communication. In the uniform-random traffic pattern, a packet is addressed to any other core with equal probability. From Fig. 4, it is observable that the performance of the WiMesh is maximum when the subnet size is 8 (i.e. number of WIs = 8). When the subnet size is lower (e.g. 4) than the optimal value, due to an increased number of WIs in the system, the interval in channel access by a WI increases. This increases the packet waiting time at the WIs and results in a lower performance than the WiMesh with subnet size of 8. Alternatively, when the subnet size is larger (e.g. 16) than this optimal value, each WI is shared by more number of cores and the traffic load at the WIs is increased. This results in a congestion at the WIs and reduces the performance of the WiMesh. As the performance for the WiMesh is maximum for a subnet size of 8 (i.e. 8 WIs), we consider this as the baseline WiMesh configuration. This configuration is also used for evaluating WiMesh with other MAC mechanisms so that consistency across all the architectures is maintained. To differentiate these WiMesh architectures with different MAC machanisms, we use the notation *WiMesh+MAC* to denote the WiMesh with a specific MAC mechanism.

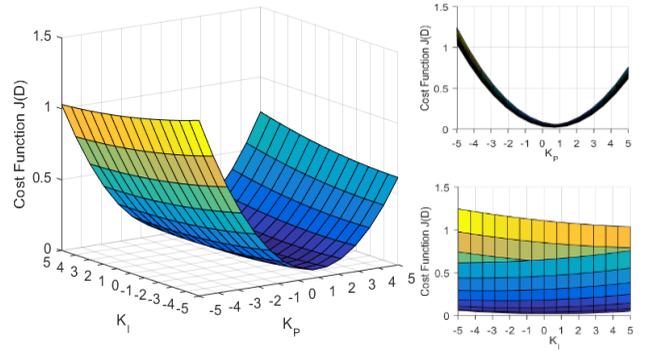

Fig. 5. Cost Function *J(D)* with respect to $K_I$ and $K_P$. Insets: only along $K_P$ and $K_I$ axes.

### 4.4 Determining the weights for Demand Prediction

In our proposed dynamic MAC mechanism, the duration of the transmission slots are allocated based on the predicted traffic demand of the WIs. To predict the traffic demand of the WIs accurately, we need to determine the values of the weights in (1) for the PID tracking, so that the prediction error is minimized. In this section, we present the two-step optimization process adopted for minimizing the prediction error. In the first step of this optimization process, we determine the optimal value for weight $K_P$ and $K_I$ by setting $K_D$ to 0. In the second step, we determine the value of the weight $K_D$ for the optimal value of $K_P$ and $K_I$. Such two-step process is commonly used for optimizing the weights in PID based control systems [39]. We select the WI with highest variation in flit arrival and determine the optimal value of the weights. This optimizes the weights for quick response to variations in traffic demands with self-similar, bursty temporal characteristics [16]. These optimal weights are then used to predict the traffic demand for all WIs. To generate the training set of actual traffic demand of the WI, we monitor the number of incoming flits at the WI per epoch for the baseline WiMesh architecture for uniform random traffic at full injection load of 1 flit per core per cycle with a self-similar injection pattern. The training set contains 5,000 traffic demand values per WI. Then, we construct a cost function $J(D)$ for capturing the root mean square (rms) prediction error, which is given by,

$$J(D) = \frac{\sum_{i=1}^{m}(D_{pred}^i - D_{Actual}^i)^2}{m} \quad (5)$$

where, $D_{pred}^i$ and $D_{Actual}^i$ is the predicted and actual traffic demand for at epoch $i$, $m$ is the total number of samples in the training set. As the cost function is based on squared value of errors, it is convex in nature and have one minima. Hence, the main goal of the optimization process is to determine the minima of (5) and the values of $K_P$, $K_I$ and $K_D$ at the minima are the optimal weights.

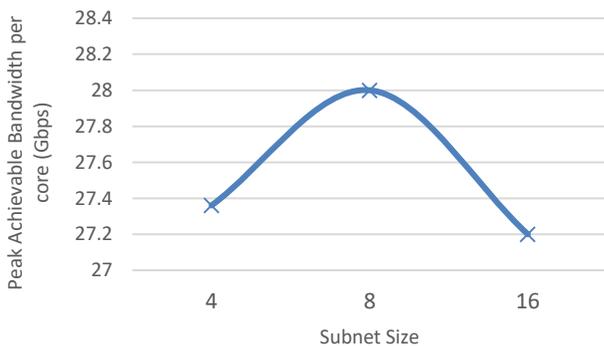

Fig. 4. Peak Achievable Bandwidth per core for the baseline WiMesh with varying subnet size.

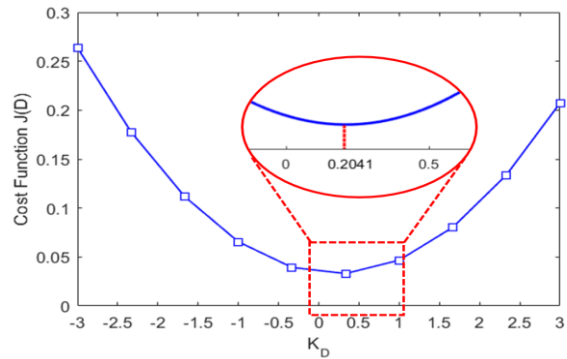

Fig. 6. Cost Function *J(D)* with respect to $K_D$.

At the first step of the two-step optimization, we set $K_D$ to 0 and use the gradient descent method [39] to find the minima of the cost function. At each iteration of the gradient descent method, the value of $K_P$ and $K_I$ is updated and the predicted traffic demand of the WIs are calculated using (1) with these updated values. This prediction is then used for determining the value of the cost function at each iteration. When the gradient descent converges to the minima of the cost function, the values of $K_P$ and $K_I$ at that iteration is considered as the optimal value. The values of the cost function with different $K_P$ and $K_I$ values at different iteration of the gradient descent algorithm is shown in Fig. 5. It can be seen from the figure that the value of the cost function is minimum when $K_P$ = 0.66 and $K_I$ = 0.13. Then, at the second step of the optimization process, the optimal $K_P$ and $K_I$ values are used to determine the value of $K_D$ using the same cost function and gradient descent algorithm used in the first step. The values of the cost function for different values of $K_D$ is shown in Fig. 6. From Fig. 6, it is observable that the cost function is minimized when $K_D$ is 0.2041. These weight values are then used in (1) to predict the demand of the WIs as required by the dynamic MAC mechanisms. Fig. 7 demonstrates the predicted traffic demand of the WIs at each epoch along with the actual traffic in that epoch. An epoch size of 100 cycles is considered here. From the figure, it is observable that the predicted demand values closely resemble the actual traffic demand and the average root mean square error (RMSE) is 1.905 flits per epoch over 20,000 cycles. We observe that due to the accurate traffic demand prediction and efficient allocation of the transmission slots, the number of wasted slots (no flits are transmitted) reduces by 1.3% and 13.5% respectively for the WiMesh+P-SAM and WiMesh+D-SAM architecture when compared to the baseline WiMesh for uniform-random traffic. We further study the effect of this reduction in the wasted slots on the performance and energy efficiency of these architectures with different traffic patterns in the next subsections.

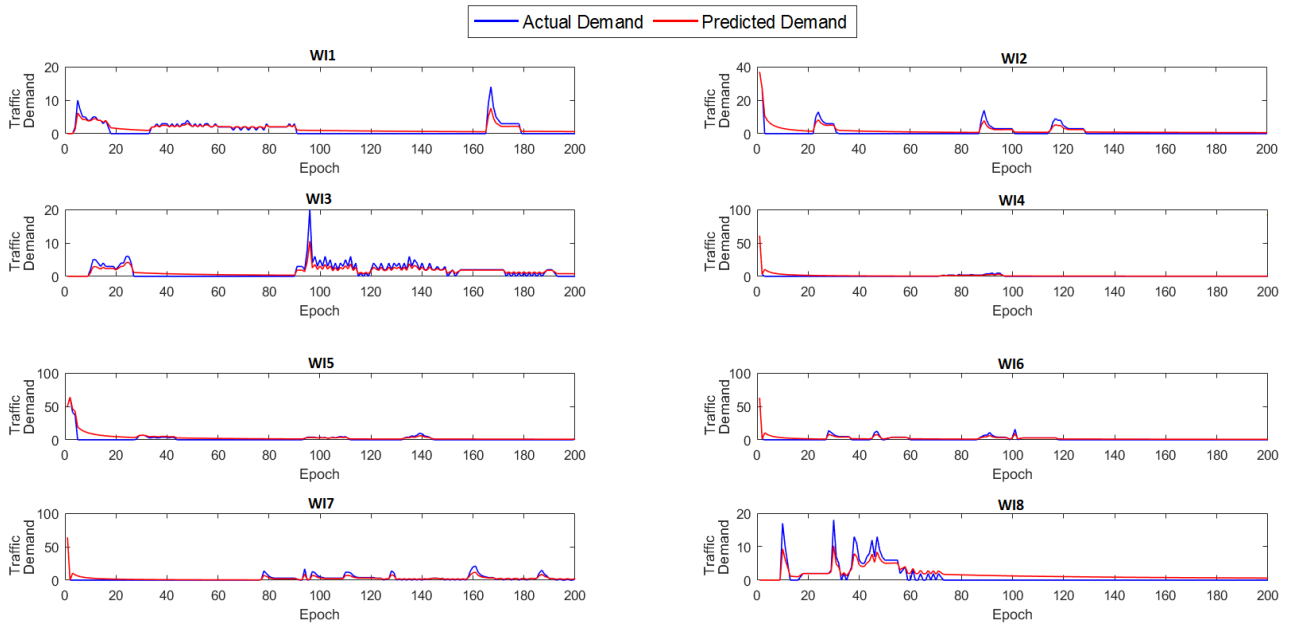

Fig. 7. Predicted traffic demand for the 8 WIs at each epoch using the PID tracking based prediction mechanism with optimal weights.

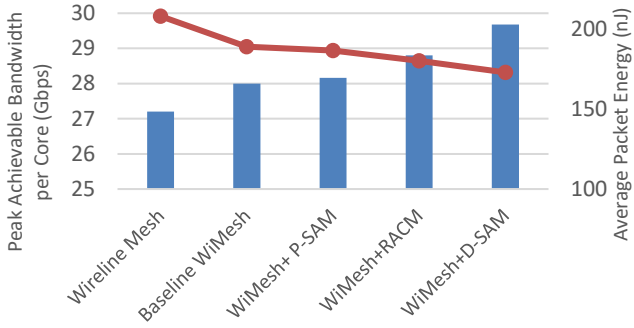

Fig. 8. Peak achievable bandwidth per core and average packet energy for uniform random traffic.

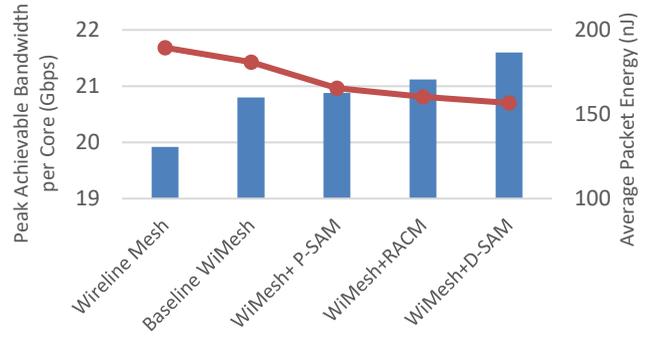

Fig. 10. Peak achievable bandwidth per core and average packet energy for hotspot traffic.

### 4.5 Performance Evaluation with Synthetic Traffic

In this section, we evaluate the performance and energy efficiency of the WiMesh architecture with different MAC mechanisms for synthetic traffic patterns. Uniform random, hotspot, and bit complement synthetic traffic patterns are used in this evaluation. A self-similar temporal injection pattern is adopted for each of the synthetic traffics in this section. We also evaluate the WiMesh architecture for broadcast traffic pattern to capture a wide range of cache coherency traffic in a multicore system.

#### 4.5.1 Evaluation with Uniform Random Traffic Pattern

The peak achievable bandwidth per core and packet energy at network saturation for both the wireline Mesh and the WiMesh architectures with different MAC mechanism for uniform-random traffic pattern is shown in Fig. 8. Due to the presence of long-range wireless shortcuts in the baseline WiMesh architecture, the peak bandwidth per core and the energy efficiency improves compared to the wireline Mesh architecture. However, in the baseline WiMesh, slots are wasted when WIs donot have a whole packet to transmit. This limits the performance gain of the baseline WiMesh. Hence, the performance of the baseline WiMesh can be improved, by allowing partial packet transmission and dynamically adjusting the transmission slots of the WIs. The dynamic WiMesh architectures studied in this paper (i.e. WiMesh+P-SAM, WiMesh+RACM, WiMesh+D-SAM), adjusts the transmission slots of the WIs at each epoch. Such allocation strategy of transmission slots enables WIs with higher traffic demand to access the wireless channel for a longer duration. This results in improved performance and energy efficiency for the dynamic

WiMesh architectures compared to the baseline WiMesh as shown in Fig. 8. However, for both WiMesh+P-SAM and WiMesh+RACM architectures, the transmission slots are adjusted keeping the duration of the epoch constant. This result in wasted transmission slots. Moreover, in the RACM, the adjustment is based on current utilization instead of predicted ones. On the other hand, in the WiMesh+D-SAM architecture the duration of the epoch is determined based on the predicted traffic demands of the WIs further reducing wasted slots. Due to the efficient allocation of the transmission slots and the adjustment in epoch, the number of slots wasted for the WiMesh+D-SAM architecture reduces by 12.4% and 10.7% when compared to the WiMesh+P-SAM and WiMesh+RACM architectures respectively. This results in improving the performance of the WiMesh+D-SAM architecture compared to the others.

The benefits of the WiMesh+D-SAM architecture is more evident in Fig. 9, where the average packet latency at different injection load is shown for the wireline Mesh and WiMesh architectures for uniform random traffic. At very low load the latency of all the architectures is the same as they have the same topology. However, at moderate load, the average packet latency for the WiMesh+D-SAM is lower than all other architectures considered in this paper. The gain in average packet latency for the WiMesh+D-SAM significantly increases at higher injection loads due to the increase in temporal and spatial variation in traffic demand. This is due to the efficient allocation of transmission slots based on the predicted traffic demand of the WIs.

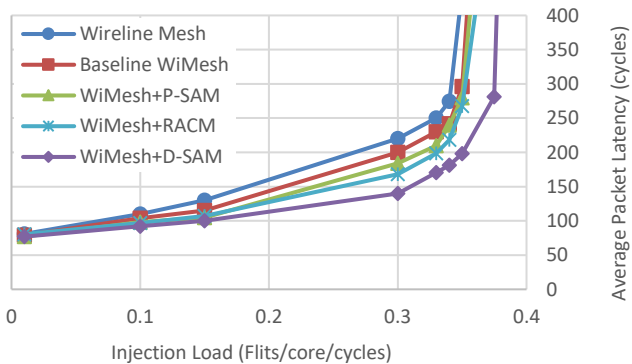

Fig. 9. Average packet latency with varying injection loads for uniform random traffic.

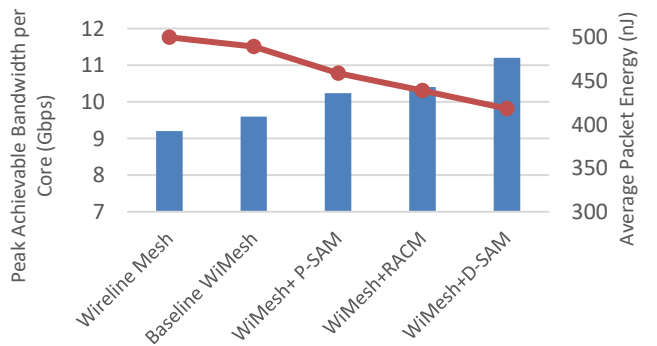

Fig. 11. Peak achievable bandwidth per core and average packet energy for bit complement traffic.

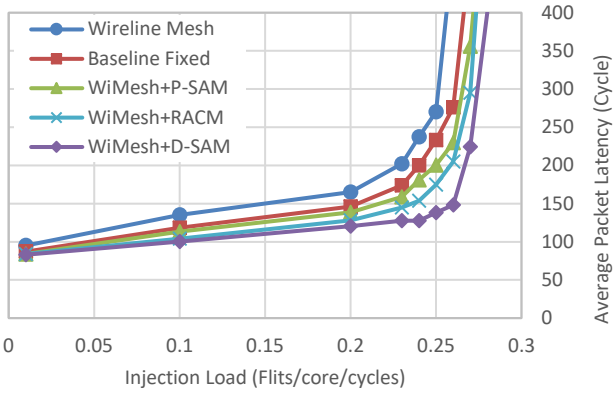

Fig. 12. Average packet latency with varying injection loads for hotspot traffic.

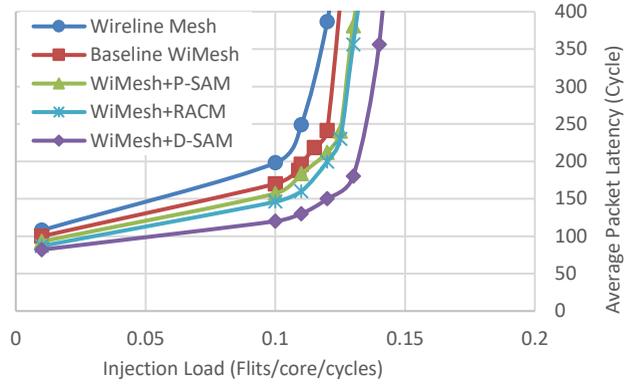

Fig. 13. Average packet latency with varying injection loads for bit complement traffic.

### 4.5.2 Evaluation with Non-Uniform Traffic Pattern

In this section, we evaluate the performance and energy efficiency of the WiMesh architectures for non-uniform synthetic traffic. Two non-uniform synthetic traffic patterns, hotspot and bit-complement are used for this evaluation. For the hotspot traffic pattern a certain volume of traffic generated from all cores is destined towards a hotspot core. All other packets are sent to other cores following a uniform distribution. This type of traffic pattern is common for directory-based cache-coherent shared memory multiprocessor system where communication among the cores and directory is frequent [40]. In our experiment, 10% of the total traffic is destined to the hotspot core which is chosen to be a core with a WI. In the bit-complement traffic pattern, packets from each core is always destined to cores whose ID is a complement of the source core. For example, packets generated from core *i* is always destined to core with ID *(N-i+1)*, where N is the number of cores in the system. Fig. 10 and Fig. 11 shows the peak achievable bandwidth per core and packet energy for the wireline Mesh and the WiMesh architectures with different MAC mechanisms at network saturation for hotspot and bit complement traffic pattern respectively. For both non-uniform traffic patterns, the baseline WiMesh architecture outperforms the wireline counterpart due to the long-range wireless shortcuts. The performance of the WiMesh architectures with dynamic MAC is better compared to the baseline WiMesh architecture. Similar to the uniform random traffic, the peak achievable bandwidth per core is highest and the packet energy is lowest for the WiMesh+D-SAM architecture among all the dynamic WiMesh architectures due to the reduction in number of wasted slots.

We also evaluate the average packet latency for different WiMesh architectures with these non-uniform traffic patterns as shown in Fig. 12 (for hotspot traffic) and Fig. 13 (for bit complement traffic). It can be seen from the figure that the average packet latency for the dynamic WiMesh architectures are lower than the baseline WiMesh at all injection load due to the reduction in number of wasted slots. The packet latency is lowest for the WiMesh+D-SAM architecture, as less number of slots are wasted due to the efficient demand based allocation. This is also consistent with our observation on average packet latency for uniform-random traffic where the packet latency is lowest for the WiMesh+D-SAM architecture.

From these evaluations of the WiMesh architectures, we see that the D-SAM MAC mechanism provides the best performance and energy efficiency due to the efficient slot allocation. Hence, for further investigation, we consider only this dynamic MAC mechanism.

### 4.5.3 Evaluation with Broadcast Traffic Pattern

Maintenance of cache coherency is a common requirement in a multicore environment [41]. As mentioned in section 4.3, a hotspot traffic pattern best represents the traffic for directory based cache coherency. However, to capture the communication pattern of broadcast based cache coherency protocols, we evaluate the proposed D-SAM MAC for broadcast traffic patterns. For this purpose, we consider a certain percentage of the traffic generated by the cores to be of broadcast nature. The rest of the traffic is unicast and the destinations are generated following the same uniform-random strategy described in subsection 4.5.1. Broadcast packets are duplicated during routing only when shortest paths to receiving cores diverge.

Fig. 14 shows the relative gain in peak bandwidth per core and reduction in average packet energy for the WiMesh+D-SAM and baseline WiMesh architectures with respect to the wireline Mesh for varying percentage of broadcast traffic. It can be observed from the figure that for different percentage of broadcast traffic the relative gain in performance and reduction in packet energy for the WiMesh+D-SAM architecture is higher than the baseline

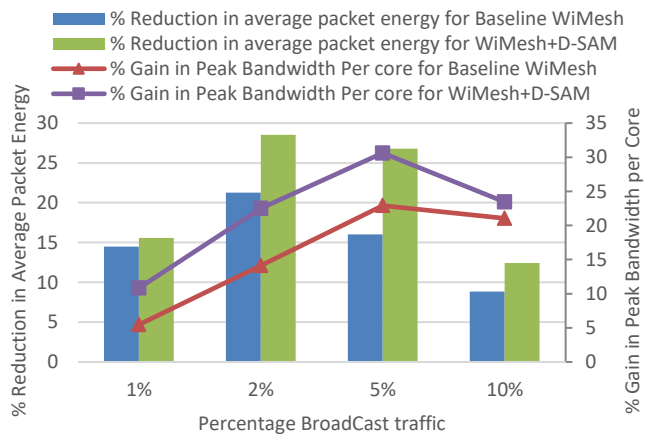

Fig. 14. Percentage gain in peak bandwidth per core and reduction in average packet energy with broadcast traffic.

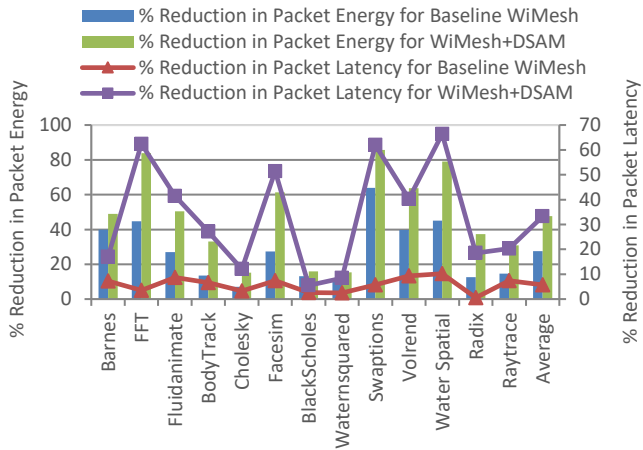

Fig. 15. Percentage reduction in packet latency and packet energy with application specific traffic.

WiMesh architecture due to the efficient allocation of the transmission slots based on the predicted traffic demands. However, the relative gain in performance and packet energy displays a convex pattern with the increase in broadcast traffic percentage for both architectures. This is because the performance is also dependent on the traffic load of the underlying Mesh architecture. Due to the inherent broadcast capability of the wireless interconnect, a single transmission via the wireless medium is sufficient for transmitting the broadcast packets to large parts of the chips directly via the WIs. The non-directional antennas and the shared wireless band makes the wireless interconnections inherently broadcast friendly. This results in the initial improvement in relative performance in both wireless architectures. However, the receiving WIs routes these broadcast packets downstream using the underlying wireline Mesh network. Increasing the percentage of broadcast packets eventually result in congestion of the WIs as they route a relatively larger number of packets compared to regular NoC switches. This eventually reduces the bandwidth and energy gain per packet with the increase in the percent of broadcast traffic as the wireless interconnection gets saturated.

### 4.6 Performance Evaluation with Application Specific Traffic

In this section, we evaluate the performance of the WiMesh+D-SAM architecture with application specific traffic patterns from PARSEC and SPLASH2 benchmark suites. To generate the application specific traffic patterns, we consider a multicore chip with 16 memory cores and 16 out-of-order (OoO) cores. Each core consists of a 32KB of L1 and 512KB of L2 cache running a Directory-Based MOESI cache coherency protocol. This core configurations are then used to extract the core-to-memory and memory-to-memory cache coherency traffic for the PARSEC and SPLASH2 benchmark applications when they are executed till completion using SynFull [15]. To map these traffic patterns to the 64-core environment, we consider 16 equal sized clusters where each cluster contains one memory core and 3 OoO cores. Three threads of the same application are then executed on the multicore chip so that each core in a cluster runs certain portion of a thread and the memory cores in the clusters are shared among the threads.

The percentage reduction in average packet latency and average packet energy for the baseline WiMesh and WiMesh+D-SAM architecture with respect to the Wireline Mesh for different application specific traffic patterns is shown in Fig. 15. The latency best represents the performance in these cases as the interconnection network is not saturated in the steady-state. The reduction in average packet latency and average packet energy for both wireless architecture varies between applications due to the variation in traffic patterns resulting in different traffic demands at the WIs. The long-range wireless shortcuts present in the baseline WiMesh architecture reduces the hop-count and provides efficient paths between core-to-memory and memory-to-memory pairs. Consequently, the energy efficiency and packet latency for the baseline WiMesh improves by 27.5% and 8.22% on an average over the wireline Mesh. On the other hand, the WiMesh+D-SAM architecture, not only provides the benefits of the long-range wireless shortcut but also enables efficient allocation of transmission slots to WIs based on their traffic demands. This dynamic adjustment of the transmission slots and the epoch in the WiMesh+D-SAM architecture enables further improment in energy efficiency and latency for the application specific compared to the baseline WiMesh. The average reduction in packet latency and packet energy for the WiMesh+D-SAM architecture compared to the Wireline Mesh is 33.29% and 47.73%. In the best case the improvements are 66.4% and 85.6% for latency and energy respectively. Hence, like synthetic traffic, the dynamic MAC enables improvement in energy efficiency and performance over the baseline T-MAC mechanism for the application specific traffic patterns.

### 4.7 Performance Evaluation with varying flit size

We discuss the performance of the WiMesh+D-SAM architecture with varying flit size and compare it with the baseline WiMesh architecture with the corresponding flit size. The increase in flit size is accommodated in the wireline network by the cost of extra hardware and silicon real estate (e.g. increasing wire width, buffer size). However, the bandwidth of the wireless interconnect is not easily scalable. Consequently, with increasing flit size, the cycles required to transmit a flit also increases. Thus, it is important to explore the effectiveness of dynamic MAC mechanism with varying flit size. Here, we investigate the performance of the WiNoC with the proposed MAC for flit sizes of 32, 64 and 128 bits. This is because as noted in [42], higher flit widths beyond 128 are shown to provide marginal gains in performance of a NoC based system. For all the cases the packet size is considered to be 64 flits.

The relative gain in peak bandwidth and reduction in packet energy for the baseline WiMesh and WiMesh+D-SAM architecture compared to the wireline Mesh with varying flit size is shown in Fig. 16. In this evaluation, we have used the uniform-random traffic pattern. From the figure, it is observable that the performance gain of the baseline WiMesh compared to the wireline Mesh diminishes with increasing flit size. This is because, the physical bandwidth of the wireless links does not increase with increase in flit size requiring more number of cylces to transmit a single flit. On the other hand, in the WiMesh+D-SAM architecture, the WI transmission slots in each epoch are dynamically adjusted based on the predicted traffic demand resulting in a lower number of wasted slots even with an increase in flit size. Due to this, the degradation in relative performance gain with increasing flit size for the WiMesh+D-SAM architecture is less steep than the baseline WiMesh. With a flit size of 128 bits the performance and packet energy of the baseline WiMesh is similar to the wireline WiMesh. On the other hand, the performance of the WiMesh+D-SAM is higher than both the baseline WiMesh and wireline Mesh for all flit sizes. The same trend is observed for reduction in packet energy.

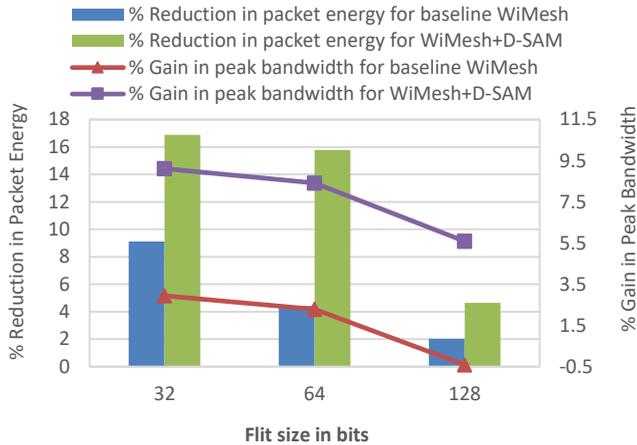

Fig. 16. Relative improvement with varying flit size for uniform random traffic

## 4.8 Evaluation with System Size

In this section, we investigate the performance and energy benefits of the dynamic MAC mechanism for larger system sizes. We considered two architectures with system size of 256 cores. In the first architecture, we considered the same number of WIs as in the 64 core system (i.e. 8 WIs). Hence, the subnet size will be much larger in this case and the WIs handle a high volume of traffic. In the second architecture, we considered the subnet size to be constant, increasing the total number of WIs in the system (i.e. 16 WIs). We adopted the same WI placemement strategy mentioned in subsection 4.3 for these architecture.

The relative gain in peak achievable bandwidth per core and packet energy for the 256 core WiMesh architectures compared to the baseline 256 core WiMesh architecture with uniform random traffic is shown in Fig. 17. It is observable from the Figure that the dynamic allocation based D-SAM MAC outperforms the baseline MAC in terms of performance and energy efficiency even for a larger system size of 256 cores irrespective of the scaling methodology. However, the improvement in performance is lower when the number of WIs are increased along with the system size (i.e. constant subnet size) compared to the case when the number of WIs remains constant. This is because, with increase in the number of WIs, the traffic is also distributed among the WIs and there is less spatial variation among the WIs. Therefore, for larger system size also the D-SAM MAC improves the performance of the WiMesh architectures compared to the baseline MAC mechanism.

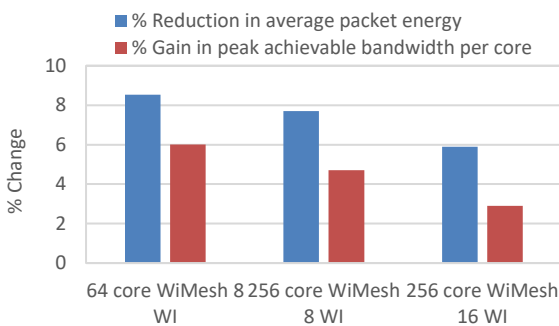

Fig. 17. Relative performance with varying system size.

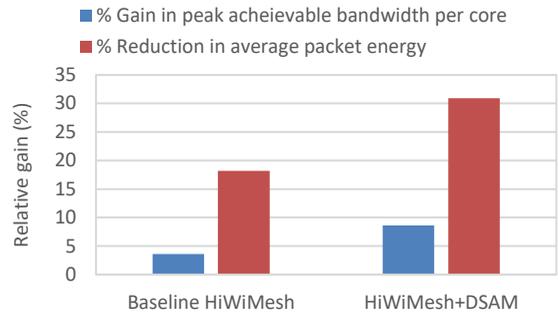

Fig. 18. Percentage gain for HiWiMesh architecture with 64 cores.

## 4.9 Evaluation with Alternative WiNoC Architecture

To justify the benefits of the dynamic MAC mechanism it is important to investigate the performance of other WiNoC architectures with the proposed dynamic MAC mechanism. In this section we study the performance and energy efficiency of an alternative WiNoC architecture equipped with the proposed dynamic MAC mechanism. For the alternative WiNoC architecture we considered a hierarchical architecture as many of the WiNoC architectures proposed in literature adopt a hierarchical approach [8], [9], [12]. Similar to the WiMesh, the underlying wireline topology of the hierarchical WiNoC is a Mesh. The Mesh is then divided into equal sized subnets and each subnet has a *hub*. The hub is a NoC switch that is connected to all other switches in a subnet using wireline links. These hubs are placed at the center of each subnet. The hubs are in turn, connected to other hubs in a Mesh fashion using wireline links. We considered the same number of subnets for the hierarchical architecture as in the WiMesh. We refer to this wireline architecture as HiMesh. To overlay the wireless links on top of this HiMesh architecture the hubs are then equipped with WIs. System-level simulations are used to find the optimal number and location of the WIs among the hubs to maximize the bandwidth of the HiWiMesh. The optimum bandwidth is observed for the configuration with 3 WIs. The HiWiMesh with the T-MAC and D-SAM MAC mechanism is referred to as baseline HiWiMesh and HiWiMesh+D-SAM architecture.

Fig. 18 shows the percentage gain in peak achievable bandwidth per core and average packet energy for the baseline HiWiMesh and the HiWiMesh+D-SAM architecture with respect to the wireline HiMesh architecture for uniform-random traffic pattern. The single-hop wireless shortcuts in the baseline HiWiMesh enables efficient data transfer between the subnets and improves the performance and energy efficiency by 3.6% and 18% compared to the wireline counterpart. On the other hand, the gain in performance and packet energy for the HiWiMesh+D-SAM architecture is higher than the baseline HiWiMesh architecture, as the transmission slots are dynamically adjusted based on the traffic demand of the WIs. This is consistent with our earlier evaluation with WiMesh architecture. Hence, the improvement in performance and energy efficiency for the D-SAM MAC can be observed in the hierarchical architecture as well.

### 4.10 Overhead Analysis

From post-synthesis analysis shown in table 1 the area of the D-SAM MAC unit is 0.0017 mm$^2$. Each transceiver occupies an area of 0.17mm$^2$ [34], [35], [36]. Therefore the overhead of the proposed MAC unit is 1% of the transceiver. The longest dimension of the zig-zag antennas is 0.3mm and they occupy a passive area using top-layer metal traces [9]. For the WiMesh architecture with 64 cores and 8WIs the total area overhead of the MAC units is only 0.0038% of the 400mm$^2$ chip.

## 5 CONCLUSIONS

Wireless interconnection is envisioned as an energy efficient communication backbone for future multicore systems. One of the key aspect for the adoption of such novel interconnect paradigm is the MAC mechanism that ensures the efficient utilization of the wireless channel based on the varying demand of the applications. In this paper, we propose the design of two dynamic MAC mechanisms that are able to efficiently adjust the transmission slots to the WIs with spatial and temporal variation of traffic demand through the WIs. Using cycle accurate simulations, we show that the prediction based dynamic MAC mechanism improves the performance of a WiNoC architecture compared to a baseline token based MAC for a wide range of synthetic and application specific traffic patterns. For application specific benchmark traffic patterns the improvements compared to a wireline mesh NoC can be as high as 66.4% and 85.6% in latency and packet energy respectively for a low area overhead of 0.0038% in a 400mm$^2$ chip. Therefore, the benefit of the proposed MAC mechanism is significant for negligible additional overhead.

### ACKNOWLEDGMENT

This work was supported in part by the US National Science Foundation (NSF) CAREER grant CNS-1553264 and grant CCF-1162123.